\documentclass[12pt]{iopart}

\usepackage{graphicx}
\usepackage{siunitx}
\usepackage[hidelinks]{hyperref}
\usepackage{sidecap}
\usepackage{orcidlink}
\definecolor{orcidlogocol}{HTML}{A6CE39}

\usepackage[sorting=none]{biblatex}
\begin{document}

    \title[Reducing suspension control noise with interferometric sensors]{Reducing suspension control noise with interferometric sensors---an experimental concept}    
    
    \author{N. L. Weickhardt$^{1}$\orcidlink{0000-0002-3923-5806}, A. Basalaev$^{2,3}$\orcidlink{0000-0001-5623-2853}, O. Gerberding$^{1}$\orcidlink{0000-0001-7740-2698}}
    
    \address{$^1$ University of Hamburg, Institute of Experimental Physics, Luruper Chaussee 149, 22761 Hamburg}
    \address{$^2$ Max Planck Institute for Gravitational Physics (Albert Einstein Institute), Callinstr.\,38, Hannover, Germany}
    \address{$^3$ Institute for Gravitational Physics of the Leibniz Universit\"at Hannover, Callinstr.\,38, Hannover, Germany}
    
    \ead{nils.weickhardt@uni-hamburg.de}

    \vspace{10pt}
    \begin{indented}
    \item[]July 2025
    \end{indented}
    
    \begin{abstract}
        One of the limiting noise sources of ground-based gravitational wave detectors at frequencies below \qty{30}{\Hz} is
        control-induced displacement noise.     
        Compact laser interferometric sensors are a prime candidate for improved local displacement sensing.
        In this paper we present the design of an experiment that aims to demonstrate the advantages of interferometric sensors over shadow sensors. 
        We mount two such interferometric sensors on two HAM Relay Triple Suspension (HRTS) systems that suspend two mirrors forming an optical cavity. As a model sensor we considered the compact balanced readout interferometer (COBRI), a sensor we are developing that is based on deep frequency modulation.
        By measuring the length stability of this cavity relative to a stable reference we aim to probe the direct motion reduction when using COBRIs for active damping and we aim to investigate their behavior and auxiliary functions, such as absolute ranging, in the context of the 6 degree-of-freedom controls of the suspensions. Here we describe the design of the experiment and simulations of the achievable noise levels that were obtained using mechanical models of the HRTS suspensions. We discuss all relevant noise sources, the modeled influence of the interferometric sensor damping and the current limitations and necessary improvements of our testing facility in terms of seismic pre-isolation to achieve a shadow sensor limited noise at around \qty{5}{\Hz}, where, according to our simulations, we can demonstrate superior performance for COBRIs in the longitudinal degree of freedom.

    \end{abstract}
    
    \section{Introduction}
        In 2015, the first direct detection~\cite{ligoscientificandvirgocollaborations2017} of a gravitational wave was achieved by the Advanced Laser Interferometer Gravitational-Wave Observatory (LIGO)~\cite{aasi2015} detectors, enabling us to observe the universe in a new way, complementary to electromagnetic radiation and astro-particles. Since then, many more gravitational waves originating from compact binary inspiraling black holes and neutron stars have been detected~\cite{theligoscientificcollaboration2021c, ligoscientificcollaboration2023a} by the LVK-Collaboration~\cite{abbott2020a}, a joint gravitational wave detector network formed by Advanced LIGO, Advanced Virgo~\cite{acernese2015} and KAGRA~\cite{akutsu2021}. In one case~\cite{abbott2017b}, a binary neutron star merger was successfully detected together with its electromagnetic counterpart, demonstrating their importance to multi-messenger astronomy.
        
        Increasing the sensitivity of these detectors at frequencies below \qty{30}{\Hz} would enable us to detect gravitational waves much earlier, from sources that are further away and with a higher signal-to-noise ratio. This would also lead to better sky localization, which is achieved through triangulation with multiple gravitational-wave detectors, aiding in the detection of particles and gamma rays that are also emitted by certain gravitational wave sources, such as merging neutron stars~\cite{abbott2017b}.
        
        Gravitational-wave detectors are the most sensitive length difference measuring devices and are based on the Michelson interferometer~\cite{aasi2015}. They measure the differential arm length (DARM) of the interferometer, influenced by space-time curvature changes induced by a passing gravitational wave, causing relative length changes which are on the order of \qty{e-21}{}. Therefore, it is necessary to isolate the interferometer optics from external forces. On Earth, one of the largest sources of noise is seismic motion, which can be decoupled from the test masses using suspensions. These suspensions provide isolation above their resonance frequencies, following an $f^{-2n}$ behavior, where $n$ denotes the number of pendulum stages, and $f$ frequency, creating quasi-free test masses at high frequencies~\cite{robertson2002}. However, the test-mass motion is amplified at the resonance frequencies, resulting in significant total root-mean-square (RMS) motion. Meanwhile, current interferometers are designed to only operate in a narrow range of motion ($\ll$ laser wavelength) where their response is approximately linear. Exceeding this range can lead to non-linear coupling and deteriorating control, and in the worst case, loss of lock and thereby reduced duty cycle, thus, an active damping system is typically used to reduce the test-mass motion at these resonances~\cite{strain2012}.
        
        Such an active damping system requires a local displacement sensor to measure the pendulum motion and an actuator to counteract that motion. This is currently achieved with shadow sensors to measure the displacement and voice-coil setups for actuation~\cite{carbone2012, akutsu2020, cooper2022}. However, since these sensors have a peak sensitivity of \qtyrange[range-phrase=--, range-units=single]{1}{0.04}{\nano \m \per \sqrt{\Hz}}, depending on the frequency range and the exact variant,
        they introduce noise into the detector, limiting its performance at frequencies below \qty{30}{Hz}~\cite{capote2025}. While control loop shaping can be used to reduce this influence in the measurement band, this in turn limits the active damping as well, leading to a compromise in the design that is dominated by the sensor noise floor.
        
        Improved local sensors with higher sensitivity would reduce control noise and decrease the RMS motion of the test masses, thereby increasing the sensitivity of the detector and its stability, as e.g. predicted for the Homodyne Quadrature Interferometer (HoQI)~\cite{cooper2018} at the future Big Beam-Splitter Suspension (BBSS)~\cite{vandongen2023}. Such sensors are not only relevant to achieve the design sensitivities of current instruments, but are even more relevant for third-generation detectors like Cosmic Explorer (CE)~\cite{reitze2019} and the Einstein Telescope (ET)~\cite{punturo2010, abac2025}. Especially the latter aims for an improvement of about 6 orders of magnitude at \qty{3}{\Hz} in comparison to current detectors, which will be nearly impossible to realize with shadow sensors for local displacement sensing.
        
        The sensor we are developing is the Compact Balanced Readout Interferometer (COBRI), which is based on ref.~\cite{gerberding2021} and employs deep frequency modulation~\cite{gerberding2015} to achieve multi-fringe sensing, absolute ranging, and a theoretical peak sensitivity of \qty{10}{\femto \m \per \sqrt{\Hz}}. In this paper, however, we simulated an experiment, which aims to show the improvement such sensors can have on the controls noise in suspensions. 

    \section{Experimental setup}

            \begin{figure}[]
            \centering
                \includegraphics[width=.6\linewidth]{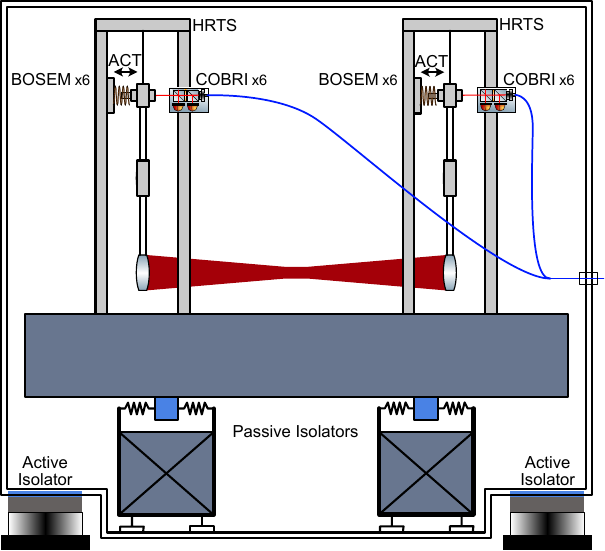}
                \caption{Sketch of the planned experimental setup. Two HRTS facing each other on a passively isolated optical table, inside VATIGrav, an actively isolated vacuum chamber. 6 DOF sensing and control at the top mass with both BOSEMs and COBRIs.}
                \label{fig:Exp_setup}
            \end{figure}
            
        To study the assumed advantages and usability of interferometric sensors in a dedicated environment, we are designing an experiment using two HAM Relay Triple Suspension (HRTS)~\cite{huddart2020a} systems, which will be used for LIGO~\cite{fritschel2014, barton2024} and ET Pathfinder~\cite{utina2022}. They were temporarily provided by the latter  to us for the experiment described in this paper. The HRTS is a triple suspension with control of all six degrees of freedom (DOFs) using Birmingham Optical Sensors and Electro-Magnetic actuators (BOSEMs)~\cite{aston} at the top mass, with a total pendulum length of \qty{390}{\milli \m}. OSEMs generally measure displacement through shadow sensing. A flag attached to the top mass partially obstructs a light beam inside the OSEM. When the top mass moves, the amount of light passing by the flag changes, providing information about its displacement. For actuation, a magnet is attached to the flag. By passing a current through the coil wrapped around the outside of the OSEM, a magnetic force is generated, which moves the flag and, consequently, the top mass. As depicted in \autoref{fig:Exp_setup}, we will mount a COBRI opposite each BOSEM, allowing us to switch from BOSEM sensing to COBRI sensing while still using the BOSEM coils for actuation, thus, allowing us to directly measure an improvement in suspension control. To reduce seismic noise, the experiment will be set up inside an actively isolated vacuum chamber (VATIGrav)~\cite{basalaev2024} on an optical table, which rests on four passive isolators.

            \begin{figure}
                \centering
                \includegraphics[width=1\linewidth]{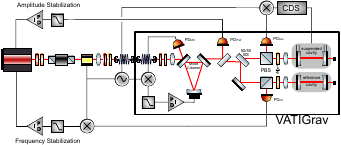}
                \caption{Proposed optical setup for the suspension cavity readout. The laser frequency stabilization is realized by locking the laser to a reference using PDH. The amplitude is stabilized via dc-offset locking with a power pick-off after the mode-cleaner in vacuum. The mode-cleaner is locked to the laser-frequency with PDH and the suspended cavity is kept on resonance with a digital control system, which also takes advantage of the PDH sidebands from locking to the reference cavity.}
                \label{fig:opt_setup}
            \end{figure}

        Measuring the improvement of control is accomplished using an optical resonator set up between the test masses of both HRTS systems, which will be read out using the Pound-Drever-Hall (PDH) technique~\cite{drever1983}, as shown in \autoref{fig:opt_setup} with design parameters in \autoref{tab:cavity_parameters}. To reduce coupling to laser frequency and amplitude noise, the laser needs to be locked to a monolithic reference cavity and stabilized in power. To minimize the effect of beam-jitter an input mode cleaner is introduced inside the vacuum chamber before coupling the laser into the cavities. Locking the mode cleaner will also be done with PDH by creating two sets of sidebands. One is being reflected by the mode cleaner to lock it, while the other set is transmitted by choosing a modulation frequency within the linewidth of the mode cleaner to lock to the reference cavity. The power stabilization will be done by introducing a pick-off behind the mode cleaner to account for power fluctuation that might be introduced by the electro optical modulator (EOM) or from the mode cleaner due to beam-jitter. Control of the suspended cavity is realized via digital system based on the LIGO Control and Data System (CDS)~\cite{ligo_scientific_collaboration_cds_2025}. 

           \begin{table}[h!]
                \caption{Design parameters of the proposed optical setup. These values are preliminary and are subject to change during the construction of the experiment.}
                \begin{indented}
                    \item[] \begin{tabular}{lcc}
                        \br
                        \textbf{Parameter}                 & \textbf{Symbol}            & \textbf{Value}    \\
                        \mr                                                       
                        Laser wavelength                   & $\lambda$                  & $\qty{1064}{\nm}$ \\
                        Laser power                        & $P_0$                      & $\qty{1}{\W}$     \\
                        Mode cleaner Finesse               & $\mathcal{F}_\textrm{MC}$  & 1000              \\
                        Mode cleaner round-trip length     & $L_\textrm{MC}$            & $\qty{53}{\cm}$   \\
                        Mode cleaner FWHM                  & -                          & $\qty{1}{\GHz}$ \\
                        Suspended cavity Finesse           & $\mathcal{F}_\textrm{sus}$ & 3140              \\
                        Suspended cavity mirror separation & $l_\textrm{sus}$           & $\qty{40}{\cm}$   \\
                        Reference cavity Finesse           & $\mathcal{F}_\textrm{ref}$ & 3140              \\
                        Reference cavity mirror separation & $l_\textrm{ref}$           & $\qty{20}{\cm}$   \\
                        \br
                    \end{tabular}
                \end{indented}
                \label{tab:cavity_parameters}
            \end{table}

        The vision goal of this experiment is to test any interferometric type displacement sensor~\cite{cooper2018, Yang2020, Smetana2022, Mitchell2024} using this setup, we, however, will focus on our COBRIs for the results shown in this paper and the initial stages of the experiment. Hence we will only refer to it in the following.

    \section{Suspension control modeling and noise discussion}
        Making assumptions about the motion of the cavity mirrors requires a set of simulation tools. One tool used is the Mathematica model for advanced LIGO suspensions~\cite{barton}, from which a state-space model can be exported that describes the mechanical properties of the suspension. This model is then read into the \texttt{spicypy}~\cite{basalaev_2023_10033638} python package, based on Python Control Systems Library~\cite{python-control2021} and \texttt{GWpy}~\cite{gwpy} packages, to compute frequency- and time-series responses of the suspension masses for various inputs, as well as to model the influence of control loops. All ASD plots presented here, unless explicitly indicated, are also calculated with \texttt{spicypy} and the Logarithmic Power Spectrum Density (LPSD) algorithm~\cite{lpsd}, originally implemented in \texttt{LTPDA} package for the LISA mission~\cite{ltpda}.
        
        In the following, we will denote transfer functions as $P$, the subscripts, e.g., on $P_{x_0,x_3}$, specify the displacement input ($x_0$: suspension point in $x$-direction) to the displacement output ($x_3$: test-mass in $x$-direction), where $x$ is the direction along the cavity axis. In the case of force inputs $F_{x_i}$, the susceptibility, i.e. the force response of the mechanical system, is given by $P_{F_{x_i}, x_i}$, where $i$ denotes one of the three masses of the pendulum suspension.
        
        The noise contribution of each sensor to the test-mass motion in the x-direction $\tilde{D}_{\textrm{S}, x_3}$ can be calculated with
            \begin{equation}
                \tilde{D}_{\textrm{S}, x_3} = \tilde{d}_\textrm{S} \cdot P_{F_{x_1}, x_3} \cdot \frac{G_{o}}{1 + G_{o}}
            \end{equation}
        where $\tilde{d}_\textrm{S}$ is the amplitude spectral density (ASD) of the sensor noise, $P_{F_{x_1}, x_3}$ is the susceptibility from a force input at $F_{x_1}$ to a displacement output at $x_3$ and $G_{o}$ the open-loop transfer function, which is given by
            \begin{equation}
                G_{o} = C_x \cdot P_{F_{x_1}, x_1}
            \end{equation}
        with the damping filter $C_x$ and the susceptibility of the top mass $P_{F_{x_1}, x_1}$.
        The damping filters we use for our analysis were published by Jeffrey Kissel in the SWG LIGO Logbook~\cite{kissel2024} and are designed to work for all DOFs on most suspensions. However, more optimal filters can be designed when interferometric sensors are used, which is an ongoing field of study~\cite{sijtsma2025} and beyond the scope of this paper. It should be noted that we plan to use more optimized filters after the initial phases of this experiment. Sensors with a higher sensitivity allow for more aggressive filtering without causing instabilities in the control loop and without introducing excess noise in the measurement band,, since less noise will couple into it. This would increase the stability of the whole system and potentially result in an improved demonstration of the benefit interferometric displacement sensors have for the observatories.  

            \begin{figure}
            \centering
                \includegraphics[width=.65\linewidth]{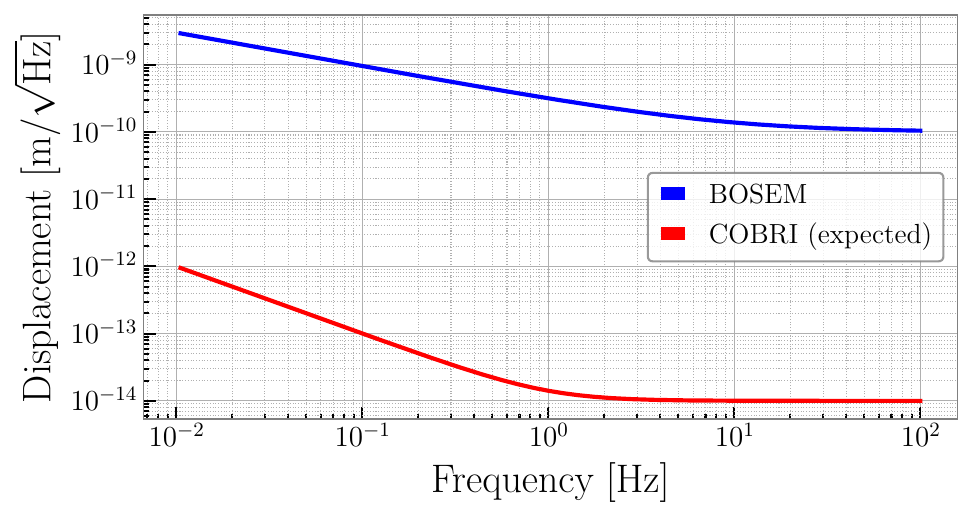}
                \caption{Amplitude spectral density models of the BOSEM- and expected COBRI sensing noise. The COBRI noise is dominated by laser freqeuncy noise at low frequencies and electronic readout noise at high frequencies.} 
                \label{fig:Sensor_noises}
            \end{figure}
            
        The sensor noise models used for the BOSEM are taken from~\cite{aston, astonb}, while the one for the COBRI is an estimation based on calculations of the Cramér-Rao lower bound~\cite{eckhardt2022}, exhibiting a rising slope of $f^{-1}$ toward lower frequencies to account for a limited suppression of laser frequency noise~\cite{Gerberding2017}. A similar noise performance using a single component DFMI-type sensor was previously achieved in ref.~\cite{Isleif2019}. Both noise curves are shown in \autoref{fig:Sensor_noises}.
        
\begin{figure}[]
                \begin{minipage}{.49\linewidth}
                    \includegraphics[width=1\linewidth]{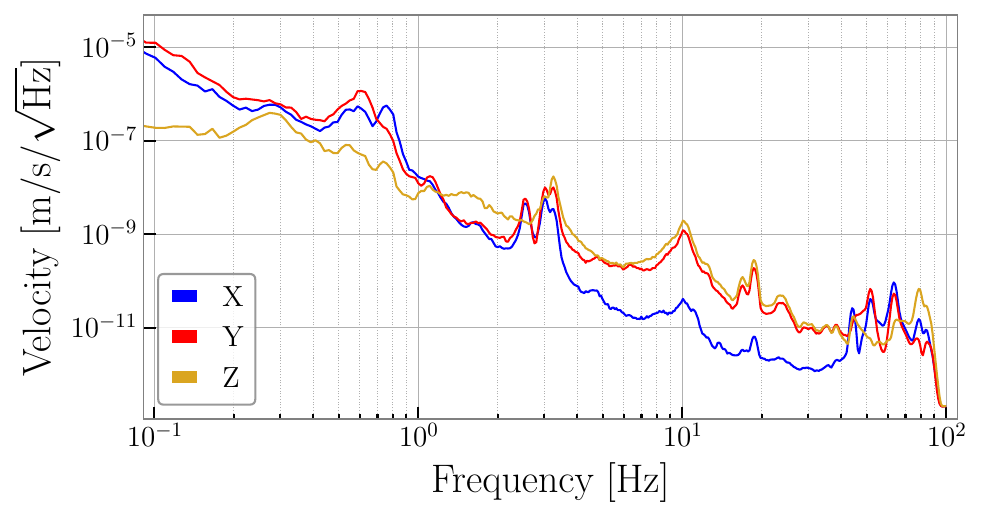}
                    \caption{\label{fig:Seis_trans}ASD of the seismic measurements in translation on the optical table with active and passive pre-isolation.}
                \end{minipage}
                    \hspace{.02\linewidth}%
                \begin{minipage}{.49\linewidth}
                    \includegraphics[width=1\linewidth]{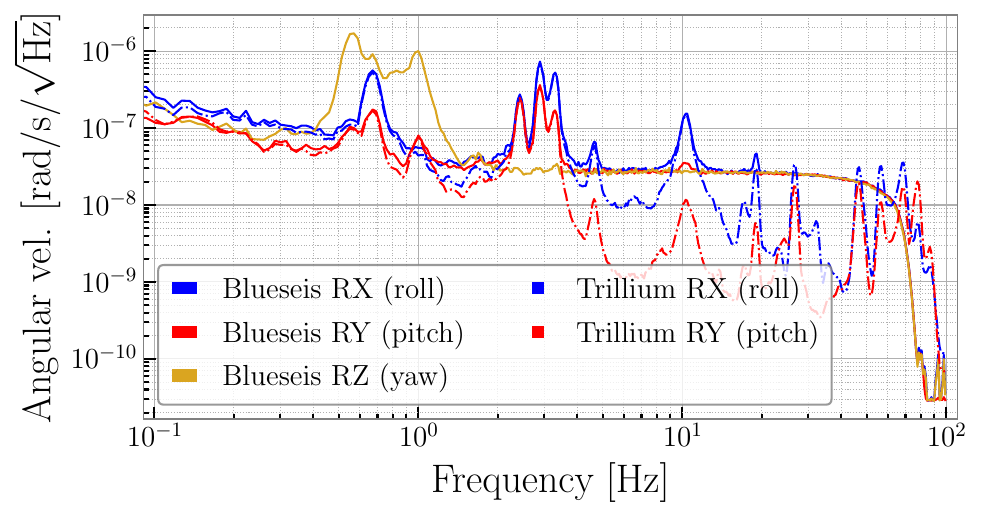}
                    \caption{\label{fig:Seis_rot}ASD of the seismic measurements in rotation on the optical table with active and passive pre-isolation.}
                \end{minipage} 
            \end{figure}

       The current seismic motion of the table inside VATIGrav was measured, with active isolation engaged, for both translations and rotations. The translational seismic motion was measured using a Trillium Horizon 120 (TH120)~\cite{nanometrics} seismometer, while rotations were measured using a Blueseis-3A~\cite{izgi2021} rotational seismometer and three TH120s set up in a triangular configuration. TH120 are conventional seismometers, relying on a proof mass to detect inertial motion, while Blueseis-3A is a fiber‐optic gyroscope. Unlike a conventional seismometer, it features flat frequency response (in velocity). The results of these measurements are shown in \autoref{fig:Seis_trans} and \autoref{fig:Seis_rot}, respectively. The rotational measurement indicates the noise floor of the Blueseis at \qty{3e-8}{\radian \per \second} for frequencies above \qty{1.5}{\Hz}, while the triangular setup of the TH120s is sensitive to signals below that. However, despite the TH120s accurately resolving the features also measured by the Blueseis-3A, caution should be exercised in interpreting these results. The frequency response of TH120s is slightly different for each individual instrument and calibration is potentially sensitive to small changes in offsets from the ideal triangle as well as precise vertical alignment of each device. Therefore, data from triangular setup of the TH120s are not used for the simulations presented here.
        
        An important source of noise for a controlled suspension is actuator noise, which we assumed is dominated by Digital-to-Analogue Converter (DAC) noise. We estimate it by assuming a white DAC noise of $\qty{5e-6}{\V \per \sqrt{\Hz}}$, for a DAC range of $\pm \qty{10}{\V}$, which corresponds to a supplied current noise to the BOSEM coils of $\tilde{I}_{d} = \qty{5e-9}{\A \per \sqrt{\Hz}}$. The actuator noise, denoted as $\tilde{D}_{a}$, is then given by 
            \begin{equation} \label{eq:Act_Noise}
                \tilde{D}_a = \tilde{I}_{d} \cdot c_{\textrm{F}} \cdot P_{F_{x_1},x_3}
            \end{equation}
        with the force constant $c_{\textrm{F}} = \qty{0.12}{\N \per \A}$~\cite{barton2024} and the susceptibility $P_{F_{x_1},x_3}$, converting the resulting force noise to a displacement of the test-mass.

        One of the fundamental limitations will be suspension thermal noise, arising from Brownian motion and described by the Fluctuation–dissipation theorem~\cite{gonzalez2000}. It was calculated using the Mathematica models. However, the calculation of the thermal noise compared to the real-world system differs in one respect; for the blade springs, the material properties of steel were used instead of stainless steel, as the required properties had not been measured and were therefore not available to us. However, since the thermal noise at high frequencies is dominated by the contribution of the last stage wires~\cite{piergiovanni2009}, this discrepancy is expected to be negligible.

        Similarly the coating thermal noise of the cavity mirrors arises from the same principle and its power spectral density (PSD) can by calculated with~\cite{Yam2015, Tait2020}
            \begin{equation}
                S_\mathrm{xx}(f) =\frac{2 k_B T}{\pi^2 f} \frac{1}{w^2} \frac{1 - \sigma_\textrm{sub} - 2 \sigma^2_\textrm{sub}}{Y_\textrm{sub}} \sum_j{b_j d_j \phi_j} 
            \end{equation}
        where $k_\textrm{B}$ is the Boltzmann constant, $T=\qty{293.15}{\K}$ the temperature, $\omega = \qty{300}{\mu \m}$ the waist radius, $\sigma_\textrm{sub} = 0.206$ and $Y_\textrm{sub} = \qty{82}{GPa}$ are the Poisson ratio and Young's modulus of the substrate (N-BK7)~\cite{schott_bk7}. $d_j$ and $\phi_j$ are the thicknesses and mechanical losses of each coating layer made of tantala ($\textrm{Ta}_2\textrm{O}_5$) and silica ($\textrm{Si}\textrm{O}_2$), which are $d_{\textrm{Ta}_2\textrm{O}_5} = \qty{236}{\nm}$, $d_{\textrm{Si}\textrm{O}_2} = \qty{347}{\nm}$, $\phi_{\textrm{Ta}_2\textrm{O}_5} = \qty{3.8e-4}{}$ and $\phi_{\textrm{Si}\textrm{O}_2} = \qty{0.4e-4}{}$. For the cavity mirrors used in this experiment we assume a 20 layer coating. $b_j$ is a weighting factor and is given by
            \begin{equation}
                b_j = \Biggl [ \frac{(1-2 \sigma_j)(1+\sigma_j)}{(1-2 \sigma_s)(1+\sigma_s)} \Biggr] \frac{1}{1-\sigma_j} \\ \times \left[ \left( 1 - n_j \frac{\partial \theta_c}{\partial \theta_j} \right)^2 \frac{Y_s}{Y_j} + \frac{(1-\sigma_s-2\sigma^2_s)}{(1+\sigma_j)^2(1-2\sigma_j)} \frac{Y_j}{Y_s} \right]
            \end{equation}
        where $\sigma_j$ and $\sigma_s$ are the Poisson's ratios of each coating and the substrate, $n_j$ the reflective index of each layer and $\partial \theta_c/\partial \theta_j$ is the sensitivity of the total coating phase $\partial \theta_c$ to fluctuations in the round-trip phase $\partial \theta_j$ in the layers. This expression can be simplified by assuming the the substrate and coating elastic layers are equal ($Y_j \rightarrow Y_s$ and $\sigma_j \rightarrow \sigma_s$), and ignoring field penetration into the coating ($\partial \theta_c/\partial \theta_j \rightarrow 0$), $b_j \rightarrow 2$ for all layers.
        
        Another fundamental noise source in the experiment is quantum noise, a combination of shot and radiation pressure noise, which arises from the power fluctuation of the laser due to shot noise. The displacement noise from shot noise is given by~\cite{black}
            \begin{equation}
                \tilde{D}_\textrm{s} = \frac{\sqrt{h c \lambda}}{8 \mathcal{F}_\textrm{sus} \sqrt{P_\mathrm{c}}}
            \end{equation}
        with a carrier power of $P_\mathrm{c}=\qty{100}{\mW}$ at $\lambda = \qty{1064}{\nm}$ and a finesse of the suspended cavity of $\mathcal{F}_\textrm{sus} = 3140$. The noise contribution, as ASD, due to radiation pressure can be calculated using
            \begin{equation}
                s_\mathrm{xx}(f) = \frac{2 \sqrt{\frac{2hcP_\textrm{c}}{\lambda}} }{c(1-R)} \cdot P_{F_{x_3}, x_3} \quad [\unit{\m \per \sqrt\Hz}]
            \end{equation}
        with the power reflectivity of the cavity mirrors $R = 0.999$ and the susceptibility $P_{F_{x_3}, x_3}$ to convert from force to displacement.  
        
        The last source of noise we took into consideration is frequency noise $\tilde{f}$ originating, ideally, only from the length fluctuations $\tilde{l}_\textrm{ref}$ of the reference cavity the laser will be locked to. It is calculated by
            \begin{equation}
                {\tilde{f}} = \frac{\tilde{l}_\textrm{ref}}{l_\textrm{ref}} \cdot f_0
            \end{equation}
        with the laser frequency $f_0$ and the reference cavity length $l_\textrm{ref}$. We assumed that the length fluctuations are dominated by a combination of coating thermal noise of the reference cavity mirrors and shot noise (radiation pressure noise effects are negligible in a monolithic cavity), as well as a temperature coupling from the ultra low expansion (ULE) glass, used as a spacer material. This thermal coupling is defined by
            \begin{equation}
                s_\mathrm{xx}(f) = \alpha \cdot s_\mathrm{TT}(f)
            \end{equation}
        where $s_\mathrm{TT}(f)$ is the temperature noise, assumed to be $\qty{1}{\mu\K \per \sqrt{\Hz}}$ at \qty{1}{\Hz}~\cite{roma-dollase2023}, and the constant of thermal expansion (CTE) $\alpha$. The CTE according to the manufacturer is $\alpha_\textrm{ULE}=\qty[separate-uncertainty = true]{0(30)e-9}{\K^{-1}}$~\cite{corning2025}, but to not underestimate the thermal coupling in our simulation we used a value of \qty{0.1}{ppm \per \K}. From here we can take the sum of the before mentioned noise contributions to calculate $\tilde{l}_\textrm{ref}$ and $\tilde{f}$ and convert to the displacement noise of the suspended cavity via
            \begin{equation}
                \tilde{l}_\textrm{sus} = \frac{\tilde{f}}{f_0} \cdot l_\textrm{sus} = \frac{\tilde{l}_\textrm{ref}}{l_\textrm{ref}} \cdot l_\textrm{sus}
            \end{equation}
        with the absolute lengths per cavity of $l_\textrm{ref} = \qty{20}{cm}$ and $l_\textrm{sus} = \qty{40}{cm}$.

    \section{Simulating noise budget for the current setup}
        For the calculation of the pendulum motion only inputs in x and pitch DOF were used and taken from the measurement of the optical table motion. Motion in all other DOFs were set to 0, as their coupling to the cavity axis motion (x-direction) is negligible~\cite{barton12079}.

        \begin{figure}
        \centering
            \includegraphics[width=.65\linewidth]{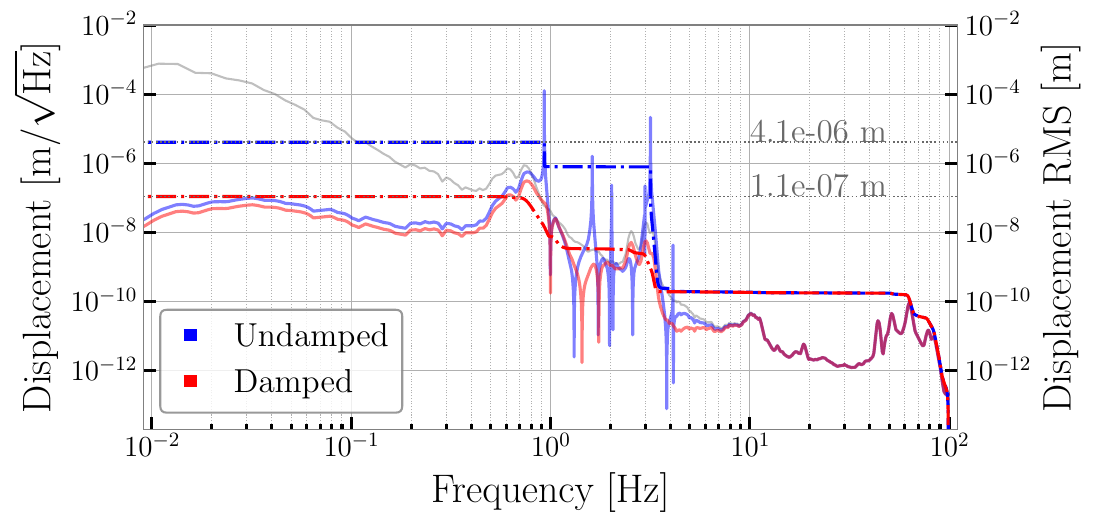}
            \caption{Expected motion spectrum (solid-lines) and RMS motion (dashed-lines) of the top mass with measured seismic motion (grey) as input.}
            \label{fig:RMS_top_mass}
        \end{figure}
            
        To estimate an appropriate DAC range, it is essential to know the expected motion at the top mass. The resulting spectrum and the RMS motion of it are presented in \autoref{fig:RMS_top_mass}. By taking the maximum RMS motion of the undamped suspension and performing the calculations from \autoref{eq:Act_Noise} in reverse, we derive a required DAC range for control of $\pm \qty{0.5}{\V}$, which includes a safety margin of a factor of two, thus, the expected noise after the coil driver is attenuated to $1/20$ of that which was estimated above.
            
        Since the property of interest in our experiment is the differential motion of the cavity mirrors, we must make an assumption about the extent to which the seismic input couples differently in the two suspensions, specifically regarding how well the mechanical properties can be matched. We assume a coupling of 1\%, meaning that 99\% of the common seismic motion is rejected. From the resulting time series of the pendulum model, the ASD was calculated using the Daniell averaging method~\cite{daniell_method}, and the contributions from other noise sources were added.

            \begin{figure}
                \centering
                \includegraphics[width=0.8 \linewidth]{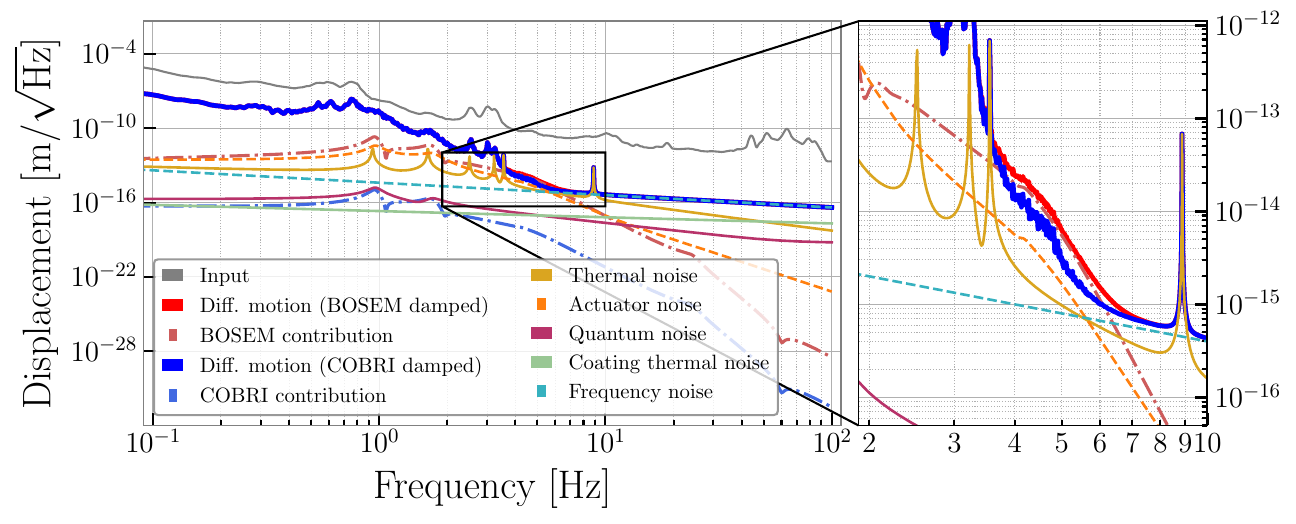}
                \caption{Noise budget of the cavity motion with the cavity constructed between the test-masses. Most important here the difference between the solid red and blue curves indicating the achievable improvement by switching to interferometric sensors.}
                \label{fig:Diff_Motion_TM}
            \end{figure}

        The noise predicted for the cavity length sensing is depicted in \autoref{fig:Diff_Motion_TM} and shows only a slight difference at frequencies from \qty{4}{\Hz} to \qty{7}{\Hz} between a damped system that uses BOSEMs (red) for sensing and one that uses COBRIs (blue). The influence of the BOSEM noise is primarily masked by the ground motion coupling, which dominates at frequencies  below \qty{7}{\Hz}, and at higher frequencies, we are limited by suspension thermal noise.
        Constructing the cavity between the penultimate masses of the suspensions would increase the coupling to sensor noise, however the coupling to ground motion is increased as well, resulting in a similarly small difference between both damping cases.

\begin{figure}[b]
            \centering
                \includegraphics[width=.65\linewidth]{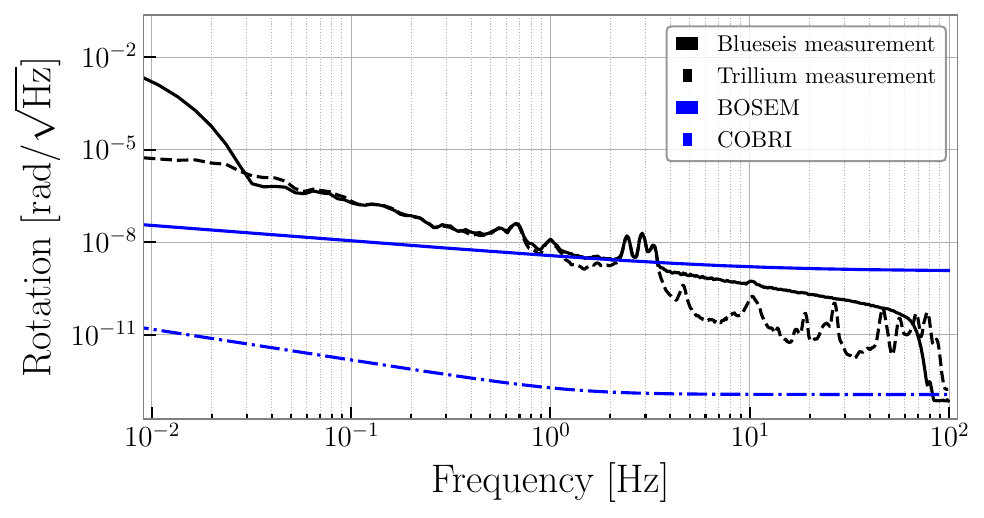}
                \caption{Measured pitch motion compared to BOSEM and COBRI pitch noise on the HRTS.}
                \label{fig:Rotation_Noise_Pitch}
            \end{figure}

        But there is also rotational motion to dampen, and in LIGO, the BOSEMs are not used to control rotations~\cite{shoemaker1996, black2010}. The reason is the large BOSEM self-noise compared to the rotational seismic motion. In our case, the BOSEMs would introduce additional rotational motion to the top mass above \qty{3}{\Hz}, as as shown in \autoref{fig:Rotation_Noise_Pitch}, which exceeds the existing rotational motion on the optical table. This makes it infeasible to use BOSEMs to dampen any rotations, not just for pitch. On the other hand, local interferometric sensors  have a much lower self-noise, which lies well beneath the measured rotational motion, making them advantageous for controlling rotations, as demonstrated in ref~\cite{Mitchell2024}.

    \section{Tightening seismic isolation requirements}
        To achieve a larger margin between a BOSEM- and a COBRI-damped pendulum, 
        the seismic isolation of the optical table needs to be improved, which would reduce the RMS motion, specifically of the top mass, thus increasing the difference of the two damping cases. The reduced RMS motion additionally allows us to further reduce the DAC range and therefore the actuator noise.

            \begin{figure}[t]
            \centering
                \includegraphics[width=.65\linewidth]{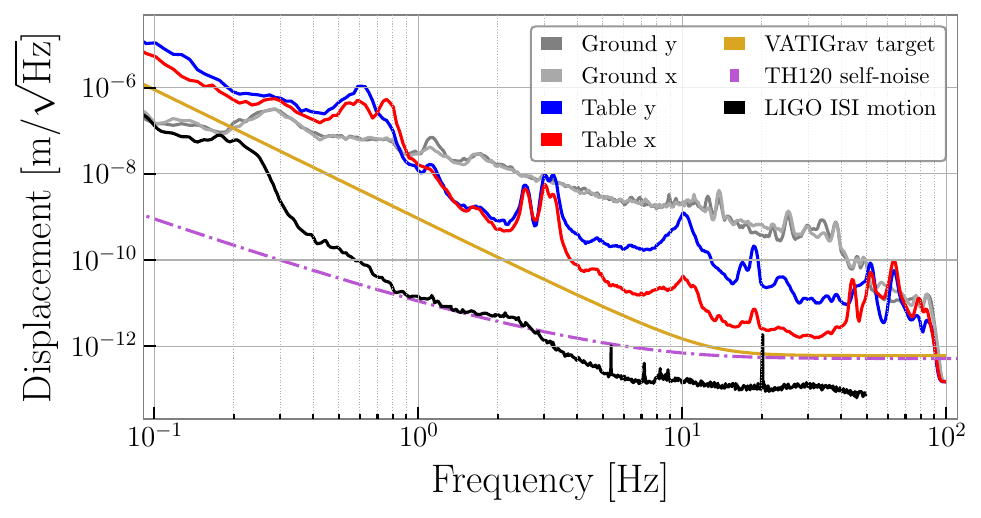}
                \caption{Target for future seismic motion on the optical table (yellow) in comparison to the current motion (blue \& red) and motion of the LIGO ISI (black)}
                \label{fig:Seismic_target}
            \end{figure}
            
        We created a target spectrum based on our current measurements, compared to the LIGO ISI~\cite{matichard2014, matichard2014a} motion at the suspension point of one of the input test masses~\cite{lantz2023} and the self-noise of the TH120~\cite{kisselsvn}. The TH120 will later be used on the optical table to create a feedback loop in conjunction with the active isolation feet of the chamber. The result, shown in \autoref{fig:Seismic_target}, consists of a $f^{-3}$ slope at low frequencies and is flat at high frequencies due to the self-noise of the TH120. Our general aim is to improve isolation by one to two orders of magnitude, depending on frequency. This positions the target spectrum between the LIGO ISI noise and our current performance.
        
            \begin{figure}[b]
            \centering
                \includegraphics[width=.65\linewidth]{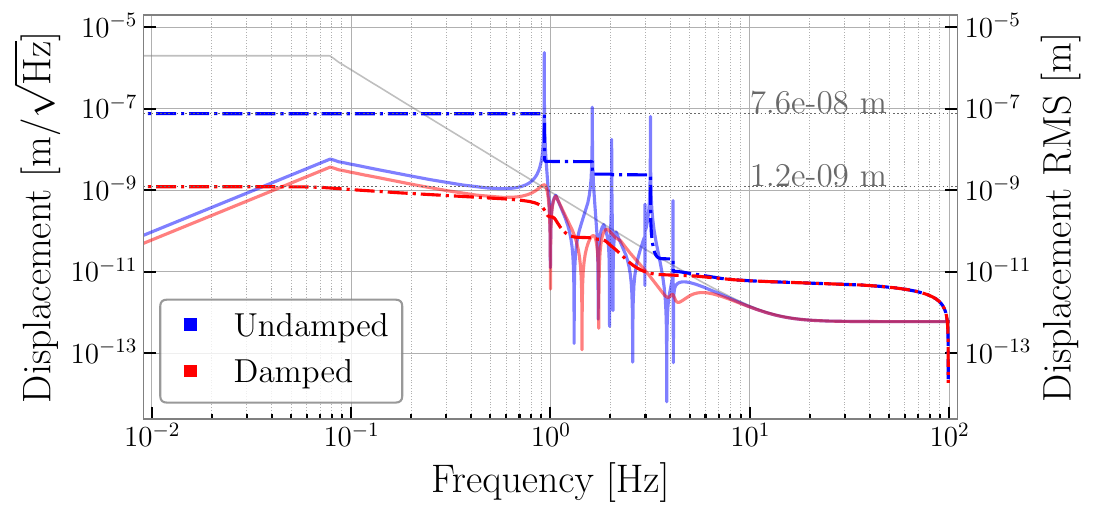}
                 \caption{Expected motion spectrum (solid-lines) and RMS motion (dashed-lines) of the top mass with seismic target (grey) as input}
                 \label{fig:RMS_top_mass_target}
            \end{figure}

        Feeding the seismic noise target into the suspension model results in a decrease in the RMS motion by roughly two orders of magnitude, as shown in \autoref{fig:RMS_top_mass_target}. This is mainly due to the absence of resonances of the passive and active isolation in the target noise, which are difficult to predict and are expected to be reduced after improving the isolation.
        This improvement would theoretically allow us to reduce the DAC range by two orders of magnitude; however, we would be limited by the noise of the Analogue-to-Digital Converters (ADC), used for BOSEM readout, at roughly $\qty{3e-7}{\V \per \sqrt{\Hz}}$. We therefore assume a conservative reduction in the DAC range by one order of magnitude.

            \begin{figure}[b]
                \centering
                \includegraphics[width=0.75\linewidth]{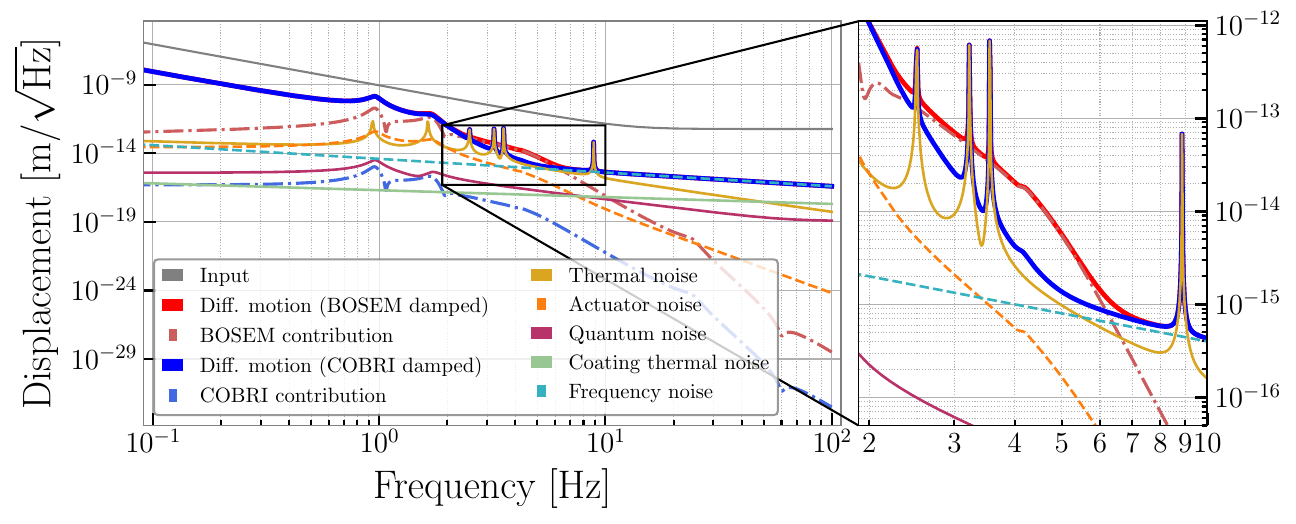}
                \caption{Target noise budget of the cavity motion with the cavity constructed between the test-masses. Most important here the difference between the solid red and blue curves indicating the achievable improvement by switching to interferometric sensors.}
                \label{fig:Diff_Motion_TM_target}
            \end{figure}

        The effect on the differential motion is depicted in \autoref{fig:Diff_Motion_TM_target}. The difference between the BOSEM- and COBRI-controlled systems increased significantly, providing a large margin to demonstrate the advantages of COBRIs over BOSEMs in terms of sensitivity and suspension damping of at least an order of magnitude.

        The seismic pre-isolation improvement that we aim for will be achieved by exchanging the inertial sensors in the commercial active isolators, with a set of three Trillium Horizon 120 that we will position on top of the vacuum chamber (for details see ref.~\cite{basalaev2024}) to enable a LIGO ISI-like active control. Additionally, the tuning of the passive dampers and the mass distribution in the passive isolation can be improved further by more careful mechanical tuning, since we do not yet achieve the nominal resonance frequencies.

    \section{Possible updates to suspension mechanics}

            \begin{figure}
            \centering
                \includegraphics[width=.65\linewidth]{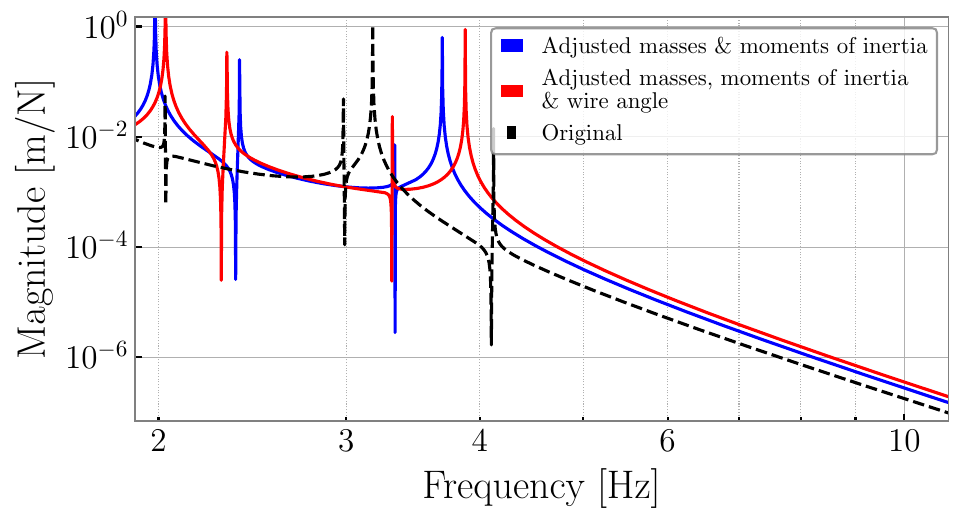}
                \caption{Susceptibility from top mass to test-mass for the original and adjusted mechanics of the HRTS.}
                \label{fig:susc_adjusted_mechanics}
            \end{figure}

        Future revisions of this experiment will feature suspensions specifically designed to test high-sensitivity interferometric local displacement sensors. These suspensions must achieve a stronger coupling between sensor noise and test-mass motion without sacrificing seismic isolation. Therefore, reducing the number of pendulum stages or actuating directly on the test mass is not feasible. As stated in~\cite{robertson2002, strain2012, plissi2000}, high coupling to the actuators can be achieved by using four wires from the mass where the actuation occurs to the test mass, by matching the masses and moments of inertia of each mass, and by tuning the angle at which the wire connects to each mass. \autoref{fig:susc_adjusted_mechanics} shows the susceptibility from the top mass to the test mass, indicating the strength of coupling between sensor noise and test-mass motion for the original system and two systems with adjusted mechanical designs. For the blue curve, matching the masses and moments of inertia resulted in an increased coupling by a factor of two over the relevant frequency range of \qtyrange[range-phrase=--, range-units=single]{3}{10}{\Hz}. By increasing the angle between the wire and the attachment points at the top, penultimate, and test masses, coupling was further increased, as visualized by the red curve. Therefore, designing a suspension with higher coupling to sensor noise should be achievable with minimal adjustments to existing designs.

    \section{Conclusion}
        In this paper we presented the setup and simulation results of an experiment, which aims to directly measure the reduction of controls noise in mirror suspensions using high sensitive interferometric local displacement sensors.
    
        We found that with the current state of our seismic pre-isolation we expect to measure a difference between a system that uses BOSEMs and one that uses the COBRIs for sensing of up to a factor of two in length stability between \qtyrange[range-phrase=--, range-units=single]{3.5}{7}{\Hz} along the cavity axis.
        By improving our seismic pre-isolation, we should be able to definitively demonstrate the improvements that our COBRIs will provide for suspension control over the currently used BOSEMs by a factor of 10. In terms of the rotational degrees of freedom (DOFs), measuring the improvement will be straightforward due to BOSEM noise dominating rotational motion in the suspension.
        Generally, this setup could be used to test any new interferometric displacement sensor, not just the COBRI.
        The final setup will feature control of all 6 DOFs with BOSEMs and COBRIs even though the focus of this manuscript was on the cavity axis. Furthermore, realizing this experiment in our environment, including the control system, we can provide a robust case for its implementation in actual observatories. While the full displacement noise improvements of local sensors will not be directly demonstrated in these experiments, they can in the future be adapted with new, more aggressive damping controllers to reduce also the RMS motion further while retaining suitable in-band noise contributions.

        The experiment could in principle benefit from another suspension design where coupling between the sensor noise and the optic would be enhanced. However, creating a new suspension for this purpose would require significant resources and research \& development. But with the HRTS we benefit from a well-tested design and a lot of existing expertise, including available mathematical models. Additionally, proving the advantages of using existing LIGO-style suspensions would underscore the improvements that our technology could bring to LIGO or future detectors.
        Nonetheless, we have shown that designing such a suspension is possible by adjusting a few key parameters of the mechanical design.

        While procurement of the major parts of the experiment has been concluded, several parts still require assembly, commissioning and, surely, noise hunting, with first major results expected by mid of 2026.

    \section*{Data availability statement}
        All data that support the findings of this study are included within the article (and any supplementary files).

    \ack
        The authors thank Mark Barton for providing the Mathematica models and our colleagues from ET-Pathfinder, especially Joris van Heijningen and Stephan Hild, for us being able to borrow the HRTSs. Furthermore we want to thank Jan-Niklas Feldhusen for assisting with the assembly of the HRTSs, Johannes Lehmann for the scientific exchange on the optical setup, the SWG (seismic working group) team from LIGO for useful discussions, Conor Mow-Lowry for his helpful comments throughout the creation of this manuscript, and Jeffrey Kissel for providing the control filters. We acknowledge funding by the Bundesministerium für Bildung und Forschung (BMBF) under project references 05A20GU5 \& 05A23GU5 and by the Deutsche Forschungsgemeinschaft (DFG, German Research  Foundation) under Germany’s Excellence Strategy—EXC 2121 “Quantum Universe”—390833306. 

    \printbibliography
    
\end{document}